\documentclass[11pt,a4paper]{article}

\usepackage{jheppub}

\usepackage{multirow, graphicx,amssymb,url,mathrsfs,amsmath}
\usepackage{wrapfig,boxedminipage,setspace,subfigure,epsfig}
\usepackage{amsxtra,amstext,latexsym,dsfont,amsfonts}
\usepackage{color,eucal}
\usepackage[dvipsnames]{xcolor}
\usepackage{float}
\usepackage{slashed,comment}
\usepackage{kotex}
\usepackage{tensor}

\usepackage{indentfirst}

\newcommand{\eg}{{\it e.g.}}
\newcommand{\ie}{{\it i.e.}}

\usepackage{appendix}

\title{Modular Hamiltonian of holographic time band states}

\author[a]{Xin-Xiang Ju,}
\author[a]{Bo-Hao Liu,}
\author[a,b]{Ya-Wen Sun,}
\author[a]{and Yang Zhao}

\emailAdd{juxinxiang21@mails.ucas.ac.cn}
\emailAdd{liubohao16@mails.ucas.ac.cn}
\emailAdd{yawen.sun@ucas.ac.cn}
\emailAdd{zhaoyang20a@mails.ucas.ac.cn}

\affiliation[a]{School of Physical Sciences, University of Chinese Academy of Sciences, Zhongguancun east road 80, Beijing 100190, China}
\affiliation[b]{Kavli Institute for Theoretical Sciences, University of Chinese Academy of Sciences, Beijing 100049, China}

\abstract{
A holographic time band is a causal incomplete boundary spacetime subregion whose causal wedge is a causal complete bulk spacetime subregion. In an AdS$_3$ spacetime with a specifically modified IR geometry, its causal wedge coincides with its entanglement wedge, which suggests the existence of a local modular Hamiltonian for the holographic time band state. In this work, we construct the local modular Hamiltonian for holographic time bands using two independent methods: from the quantum information properties of the time band state and from the construction of consistent geometric modular flows. Both methods lead to the same unique result of the local modular
Hamiltonian, reflecting the intrinsic property of the time band state. The entanglement
first law has also been checked to hold for the simplest time band state. This is a substantial addition to the known holographic subsystems with a local modular Hamiltonian, beyond the few cases previously identified.
}

\begin{document}
\maketitle
\newpage
\section{Introduction}

\noindent In AdS/CFT correspondence \cite{Maldacena:1997re}, it has been found that a boundary subregion is dual to its entanglement wedge in the bulk, \ie, the causal domain of the spatial subregion enclosed by the Ryu-Takayanagi surface \cite{Ryu:2006bv,Ryu:2006ef}, which is the subregion-subregion duality \cite{Czech:2012bh,Wall:2012uf,Headrick:2014cta,Bousso:2022hlz,Espindola:2018ozt,Saraswat:2020zzf,dong2016reconstruction,Harlow:2018fse,Bousso:2012sj}. This has been generalized to the subalgebra-subregion duality \cite{Leutheusser:2022bgi} which extends the entanglement wedge to general causal complete bulk spacetime subregions enclosed by convex surfaces by constructing a Type III von Neumann subalgebra. This bulk subregion could also be understood as the gravitaional subregion that well-defined accelerating observers could observe, namely the hole-ography in \cite{Balasubramanian:2013rqa,Balasubramanian:2013lsa,Hubeny_2014,Ju:2023bjl,Ju:2023dzo} and the convexity condition of the enclosing surface translates into the condition that worldlines of bulk accelerating observers do not intersect.
The corresponding subalgebra is associated with a boundary causal incomplete spacetime subregion, {a time band where observers living inside can only have a finite lifetime.} The edges of this time band are two ``time cutoff" spacelike hypersurfaces with arbitrary shape. The time band itself is the intersection of the bulk spacetime subregion and the conformal boundary.

One interpretation of the time band state is that it corresponds to a reconstructed global state from local density matrices of subregions whose causal diamonds are within the time band \cite{Czech:2014tva} and its {von Neumann} entropy is given by the {area of the convex surface in the bulk, evaluated by the differential entropy formula} \cite{Balasubramanian:2018uus,Hubeny:2014qwa,Headrick:2014eia,Myers:2014jia,Czech:2014wka}. The differential entropy is generally larger than the original entanglement entropy of the spatial subregion due to {the fact that the area of the RT surface is always smaller than the convex surface area} \cite{Hayden:2004wwj,Fawzi_2015}. {However, whether this quantum state exists is questioned in \cite{Czech:2014tva}.} A new approach to analyzing the time band state is to embed the bulk spacetime subregion into the geometry with a specific modification in the IR regime \cite{Ju:2024xcn}. Under this construction, the bulk spacetime subregion becomes an entanglement wedge, and the time band state, which corresponds to the modified bulk geometry, can be obtained from the subregion-subregion duality. This confirms the existence of the time band state, in contrast to the claim of its non-existence in \cite{Czech:2014tva}.

In the IR modified geometry, the entanglement wedge of the time band {coincides} with its causal wedge, and this strongly suggests that the time band state should have a local modular Hamiltonian \cite{Chen:2022eyi,Leutheusser:2022bgi}. Modular Hamiltonian {plays an important role in the study of holographic entanglement entropy and quantum information theory}. It is an operator defined by {the logarithm of the reduced density matrix} \cite{Haag1993LocalQP}. It is important in the definition of the relative entropy \cite{Araki:1976zv,Casini:2008cr,Jafferis:2015del,Blanco:2013joa}, which could be used to distinguish different quantum states. In holography, the modular Hamiltonian is also employed to prove the RT formula \cite{Ryu:2006bv, Ryu:2006ef, Hubeny:2007xt} for the entanglement entropy in certain simple cases \cite{Casini:2011kv}. As an operator, modular Hamiltonian is usually 
highly nonlocal, which cannot be written as an integral of local operators. Only in certain simple cases, people have found local modular Hamiltonians, including the modular Hamiltonian for the Rindler wedge \cite{Bisognano-Wichmann,Casini:2008cr} and for spherical spatial subregions in a conformal field theory \cite{Hislop:1981uh,Casini:2011kv}. 

A local modular Hamiltonian is essential in lifting the Type III von Neumann subalgebra in a generalized gravitational spacetime subregion to a Type II subalgebra \cite{Jensen:2023yxy} in order to define an entropy for the subsystem, utilizing an observer with a Hamiltonian following \cite{Chandrasekaran:2022cip}. It has been conjectured in \cite{Jensen:2023yxy} that for general gravitational subregions that are not entanglement wedges, a local modular Hamiltonian could be always associated by picking a vector $\xi$ which is Killing near the enclosing surface of the subregion, as the enclosing surface looks locally like Rindler space at short enough distances. This vector $\xi$ looks like a boost with constant surface gravity near the entangling surface and it was conjectured in \cite{Jensen:2023yxy} to generate a modular flow for some state of this subregion. 
 
  In this work, we obtain the modular Hamiltonian for the general holographic time band state in AdS$_3$/CFT$_2$, which is expected to be local {as its causal wedge coincides with its entanglement wedge.} According to the conjecture in \cite{Jensen:2023yxy}, we could always define local modular Hamiltonians from the bulk vector $\xi$ which is Killing at the enclosing surface. However, there are infinitely many possibilities for $\xi$. It is not known if one of them and which one of them corresponds to the time band state with a particular entanglement structure. We will show that a unique form of the modular Hamiltonian could be obtained from constraints imposed by the quantum information properties of the time band state and by constructing consistent geometric modular flows using an observer language. Remarkably, these two independent methods lead to identical results. The result is also consistent with a bulk construction of \cite{Jensen:2023yxy}. {This work provides a new addition to known subsystems which possess a local modular Hamiltonian. It is also an explicit generalization to known geometric modular flows which is not generated by Killing vectors in the bulk or conformal Killing vectors at the boundary.} {The generalizaton of these calculations to higher dimensions will also bring new insights into the causal holographic information problem \cite{Bousso:2012sj,Hubeny:2012wa,Kelly_2014}.}

\section{Holographic time band states in IR modified geometry}

\noindent We review the holographic time band state in a specific IR modified geometry so that the bulk dual spacetime subregion of the time band state becomes an entanglement wedge in the new geometry while the geometry inside the subregion is not affected.

As a bulk {causal wedge (CW)} which corresponds to the boundary time band is in general not an entanglement wedge (EW), the boundary state cannot be studied from the subregion-subregion duality. In \cite{Ju:2023bjl, Ju:2024xcn}, the IR geometry was modified into a specific extremal geometry and {CWs} become {EWs}. Specifically, given an arbitrary convex IR region on a Cauchy slice, one can modify the geometry inside the IR region by adding matter inside {and on the edge of} the IR region, and evolve this Cauchy slice in both directions of time so that we can get the whole bulk spacetime. In the spherical extremal case in \cite{Ju:2024xcn}, the IR region is replaced by a hemisphere, which is an entanglement shadow where no RT surfaces can penetrate. Instead, the RT surfaces can only wrap around the IR region, becoming convex curves from the viewpoint of the UV geometry. In this case, one can find that the entanglement wedge of a boundary subregion coincides with its causal wedge, {both being the original bulk subregion corresponding to the time band state} \cite{Ju:2024xcn}. This is strong evidence, implying a local form of the modular Hamiltonian of this boundary subregion without {any known} exceptions {for now}. Note that in this paper, we focus on a Cauchy slice with time reflection symmetry.

Moreover, in the language of von Neumann subalgebra, as the bulk entanglement wedge $R$ in the modified IR geometry coincides with the causal wedge in pure AdS, the subalgebra $\mathcal{X}_{R}$ on the boundary dual to the modified IR geometry is equivalent to the ``Y algebra" $\mathcal{Y}_{R}$ in pure AdS\footnote{It is worth noting that the complement subalgebra of $\mathcal{X}_{R}$ in the modified IR geometry is no longer a ``Y algebra" in pure AdS; a specific discussion will be presented in a forthcoming paper \cite{Ju:2025}.} \cite{Leutheusser:2022bgi}.

All these facts confirm the existence of the dual time band state. {It} has a specific quantum information property of vanishing conditional mutual information for subregions that have no causal connection in the time band and has local density matrices determined by subregions whose causal diamonds are fully within the time band. This contradicts the result in \cite{Czech:2014tva} that a time band state does not exist and we will show in Section \ref{sec3} that a flaw in their argument invalidates their conclusion and the modular Hamiltonian obtained therein is not the correct one for the time band.

\section{Modular Hamiltonian from quantum information properties of the time band}\label{sec3}

\noindent In this and the next section, we derive a unique local modular Hamiltonian for the time band states in two distinctive ways: from the quantum information properties of the time band state and from construction of geometric modular flows using observer worldlines, both leading to the same unique form of local modular Hamiltonian. 

\subsection{The vanishing CMI condition}

\begin{figure}[h]
\centering
     \includegraphics[width=12cm]{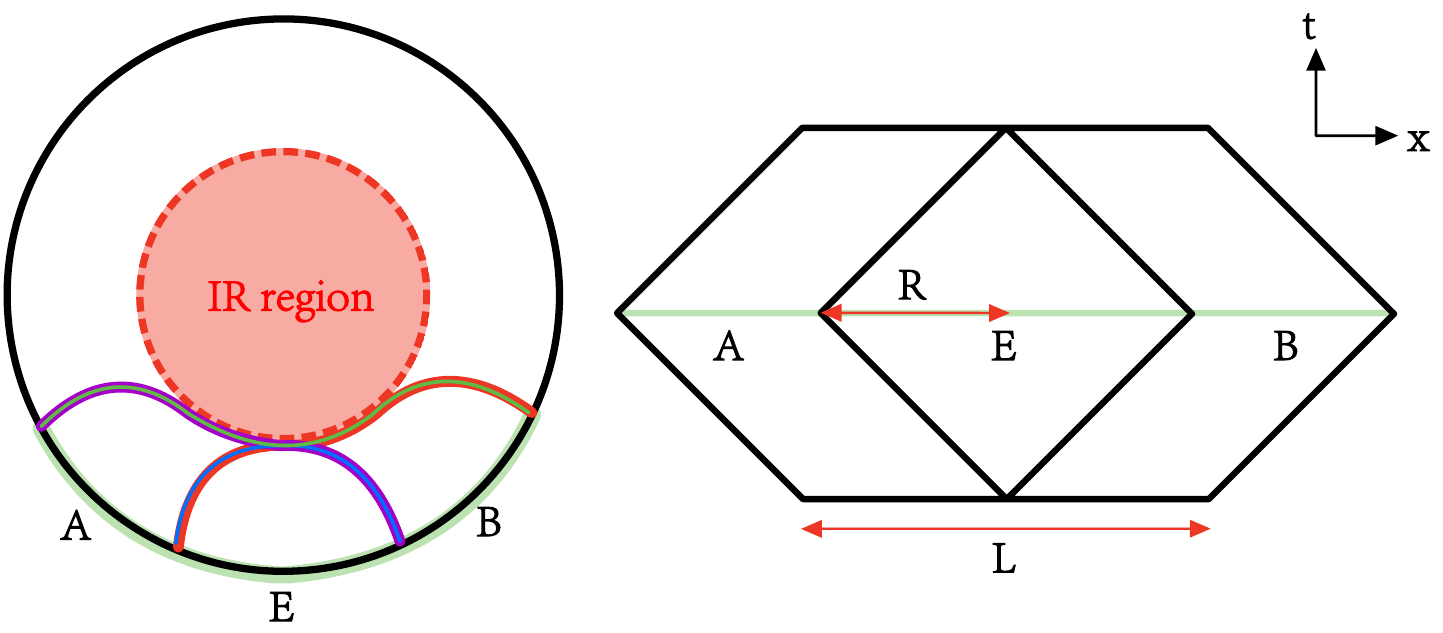}
 \caption{Left: RT surfaces corresponding to the subsystems $AE$, $BE$, $E$ and $ABE$ in the time band state in {the IR modified geometry of} AdS$_3$. Right: the time band on the boundary. $E$'s causal diamond touches the edge of the time band while $A$ and $B$ are causally disconnected in the time band with vanishing conditional mutual information.}\label{F1}
\end{figure}

\noindent In \cite{Ju:2024xcn}, we show that on the IR modified geometry, states $\tilde\rho$ associated with four {subregions} {$AE$, $BE$, $E$, and $ABE$} (distinguished from corresponding subsystems $\rho$ in the vacuum state of the original AdS$_3$ geometry) with partially coincided RT surfaces as shown in Figure \ref{F1} satisfy
\begin{equation}\label{cmivan}
    I(A:B|E) \equiv S(\tilde{\rho}_{AE}) + S(\tilde{\rho}_{BE}) - S(\tilde{\rho}_{E}) - S(\tilde{\rho}_{ABE}) = 0,
\end{equation} where $E$ should be large enough so that its RT surface at least touches the edge of the IR region as shown in Figure \ref{F1}.
In quantum information theory, (\ref{cmivan}) is equivalent to \cite{Hayden_2004,Ruskai_2002,Lieb:1973cp}
\begin{equation}\label{KCMI}
\begin{aligned}
    \log \tilde{\rho}_{AEB} &= \log \tilde{\rho}_{AE} + \log \tilde{\rho}_{EB} - \log \tilde{\rho}_{E},\\
    \Rightarrow K_{AEB} &= K_{AE} + K_{EB} - K_{E},
\end{aligned}
\end{equation}
where $K$ denotes {their} corresponding modular Hamiltonian. 

In \cite{Czech:2014tva}, the authors used this formula to argue that the von Neumann entropy of the boundary state cannot be as large as the differential entropy, thereby concluding that the holographic time band state here does not exist. However, in their derivation, they mistakenly took \(\tilde{\rho}_{AE}\) and \(\tilde{\rho}_{BE}\) to be density matrices reduced from the vacuum state. They then claimed that the corresponding CMI could still be treated as zero, which is a second-order infinitesimal term when \(A\) and \(B\) are infinitesimally small.

In fact, \(\tilde{\rho}_{AE}\) and \(\tilde{\rho}_{BE}\) should be the time band states corresponding to the modified bulk geometry, which cannot be reduced from the vacuum state. As a result, for this modified geometry, CMI is exactly zero even for finite $A$ and $B$. If, instead, one assumes these density matrices are reduced from a vacuum state, CMI is indeed a second-order term but is not negligible, because any other state will also have a second-order infinitesimal CMI for infinitely small \(A\) and \(B\). Crucially, such states do not satisfy \eqref{KCMI}. Consequently, the conclusion in \cite{Czech:2014tva} about the non-existence of the time band state does not hold. We will provide a more detailed explanation in a forthcoming work \cite{Ju:2025}.

\subsection{Other properties}\noindent
As the entanglement wedge and causal wedge of the time band state coincide in the IR modified geometry, this strongly suggests that the modular Hamiltonian associated with the time band state {is local}. We assume that the local modular Hamiltonian of the time band state has the following form
    \begin{equation}\label{LMH}
        K_{tb} = \int f_{T}(x) T_{00}(x) \, dx,
    \end{equation} where $f_T$ is regarded as the entanglement temperature.
 The task is to {determine} the exact form of $f_T(x)$. We first consider the simplest time band which has a uniform time cutoff in the whole spatial subregion and the bulk IR edge is parallel to the boundary in Poincaré coordinates. 
To determine the exact form of the modular Hamiltonian in this case, only formula (\ref{KCMI}) is not enough. We also need some other natural constraints as follows. 
\begin{itemize}
    \item I. $f_T(x)$ must be a continuous even function.
    \item II. When $L\to 0$, we expect $f_T(x) \to f_R(x) = \pi \frac{R^2 - x^2}{R}$.
    \item III. When $L\to \infty$, $ f_T(x) \leq \pi R$.
    \item IV. We expect the entanglement temperature of a larger time band to be always greater than that of its sub-time band.
\end{itemize}
Let us explain them one by one. The first two conditions are natural, which make the time band state consistent with known results. The third condition comes from the non-negativeness of the relative entropy, as found in \cite{Czech:2014tva}. The fourth condition is known valid in the special cases where those time bands are causal diamonds, and we will provide a boundary observer explanation for it in the next section.

    \subsection{Resulting formula of local modular Hamiltonian}
    
    \noindent In Appendix A, we proved that for a time band spanning from $-L/2-R$ to $L/2+R$ with a uniform spacelike boundary, the only function $f_T(x)$ which satisfies those constraints and formula (\ref{KCMI}) is
\begin{equation}\label{flat}
    f_T(x) = \max\{\tilde{f}_{R_i}(x)\} = \left\{
    \begin{aligned}
        &\pi\frac{R^2 - (x + L/2)^2}{R} \quad\quad\quad (x \in [-L/2 - R, -L/2]) \\
        &\pi R \quad\quad\quad\quad\quad\quad\quad\quad\quad (x \in [-L/2, L/2]) \\
        &\pi\frac{R^2 - (x - L/2)^2}{R} \quad\quad\quad (x \in [L/2, L/2 + R]), \\
    \end{aligned}
    \right.
\end{equation}
where $R_i$ represents all possible causal diamonds fully inside this time band with time reflection symmetry, with $R_i$ being its size, and $\tilde{f}_{R_i}(x) =\pi \frac{R_i^2 - (x - x_i)^2}{R_i}, (x_i-R_i<x<x_i+R_i)$ is the entanglement temperature of reduced substates living in $R_i$. {The maximum is taken over the values for all these reduced density matrices $R_i$ containing $x$, as shown in Figure \ref{hyperconsistent}.}

This is a very interesting result, because the function of entanglement temperature $f_T(x)$ of the time band state behaves very similarly to the entanglement wedge nesting property, as a result of the vanishing CMI condition in both the physics of $f_T(x)$ and the RT surface as explained in the follows. In fact, the RT surface of a time band state in the IR modified geometry is the envelope surface of all its substates whose RT surface touches the edge of the IR region, which results in the vanishing CMI condition in formula (\ref{cmivan}). At the same time, the maximizing formula of $f_T(x)$ in (\ref{flat}) indicates that $f_T(x)$ is an envelop function of all the $\tilde{f}_{R_i}$ functions of its substates whose RT surface touches the edge of the IR region which results in formula (\ref{KCMI}). As (\ref{KCMI}) is equivalent to (\ref{cmivan}), this means that the behavior of $f_T(x)$ is closely related to the nesting property of the entanglement wedge and both of these two come from the vanishing CMI condition. 
Due to these facts, we propose a natural generalization that the entanglement temperature of a more general time band state is also the maximum of the entanglement temperature of all its substates whose RT surface touches the edge of the IR region, and one can easily verify that (\ref{KCMI}) holds with this $f_T$ function. {With this rule, the modular Hamiltonian could be obtained for any given time band state, and its final form will depend on the shape of the time band.}

\section{Modular Hamiltonian from geometric modular flows}\noindent

\noindent Aside from the derivation of the modular Hamiltonian from the constraint of the vanishing of CMI in the time band state, in this section we provide another method to obtain the local modular Hamiltonian for the time band state by constructing geometric modular flows utilizing the boundary observers in the time band and we will show that {this method agrees with the result from the previous section}, leading to the same form of the modular Hamiltonian.

\subsection{Geometric modular flows from worldline of observers}
\noindent {As conjectured in \cite{Jensen:2023yxy}, a bulk modular flow generated by the vector $\xi$ could induce a boundary modular flow. From the bulk, we could see that $\xi$ is essentially the tangential vector of worldlines of bulk accelerating observers in the subregion with the enclosing surface being their horizon. The tangential vector of corresponding boundary observer worldlines would generate the boundary modular flow and the worldlines are essentially the trajectories of the geometric modular flow. Under the constraint that $\xi$ has to look like a Killing vector near the enclosing surface, there are still infinitely many possibilities for the observer worldlines. Therefore, we need to construct appropriate worldlines of boundary observers, \ie, the geometric modular flows.  }

 The time band at the boundary is a causal incomplete subregion with a spacelike boundary. In this causal incomplete subregion, the observers {generating the modular flows} could only last a finite time to observe \cite{Czech:2014tva, Balasubramanian:2013lsa}. Due to causality constraints, observers could only detect the entanglement structures within their causal domain, while the long-range entanglement beyond their reach is undetectable and eliminated. Degrees of freedom whose entanglement is removed are instead entangled with an outside system from the perspective of these observers. We found that this picture \cite{Ju:2023bjl, Ju:2023dzo, Ju:2024xcn} provides the right entanglement entropy of the time band state, which is equal to the area of the dual convex curve in the bulk, \ie, the RT surface of the time band state in the IR modified geometry. This long range entanglement elimination is consistent with the vanishing CMI property of the time band state. 

We propose two natural conditions that uniquely determine the set of worldlines of observers {which generate the geometric modular flow} as follows.
\begin{itemize}
\item I. The worldlines of these observers must be orthogonal to the edge of the time band.
\item II. The worldlines of these observers must be timelike hyperbolas with time reflection symmetry.
\end{itemize}
The first condition comes from the requirement that the worldlines have to be determined by the shape of the time band, which is crucial in determining the quantum entanglement structure of the state. Therefore, we have to impose this condition to incorporate the effect of the shape of the time band. The second condition is motivated by the modular flows in the causal diamond. As we expect that the time band state has the same behaviour as the causal diamond in vacuum with respect to those observers, we naturally demand that the worldlines of those time band observers to be the same as in the causal diamond, \ie, {timelike hyperbolas}. Under these two conditions, the worldlines would finally be uniquely and naturally determined. 
Note that we must guarantee that the worldlines of these observers never intersect to make those observers physically consistent. Fortunately, we find that as long as the time band is generated by a convex curve in the bulk, this is always true. The specific geometric proof can be found in Appendix B. {Overall, as shown in Figure \ref{hyperconsistent}, we can always naturally construct non-intersecting hyperbolas serving as the geometric modular flows for an arbitrary time band, determined from those two conditions.}

\subsection{Modular Hamiltonian derived by consistent geometric modular flows}
\begin{figure}[H]
\centering
     \includegraphics[width=10.5cm]{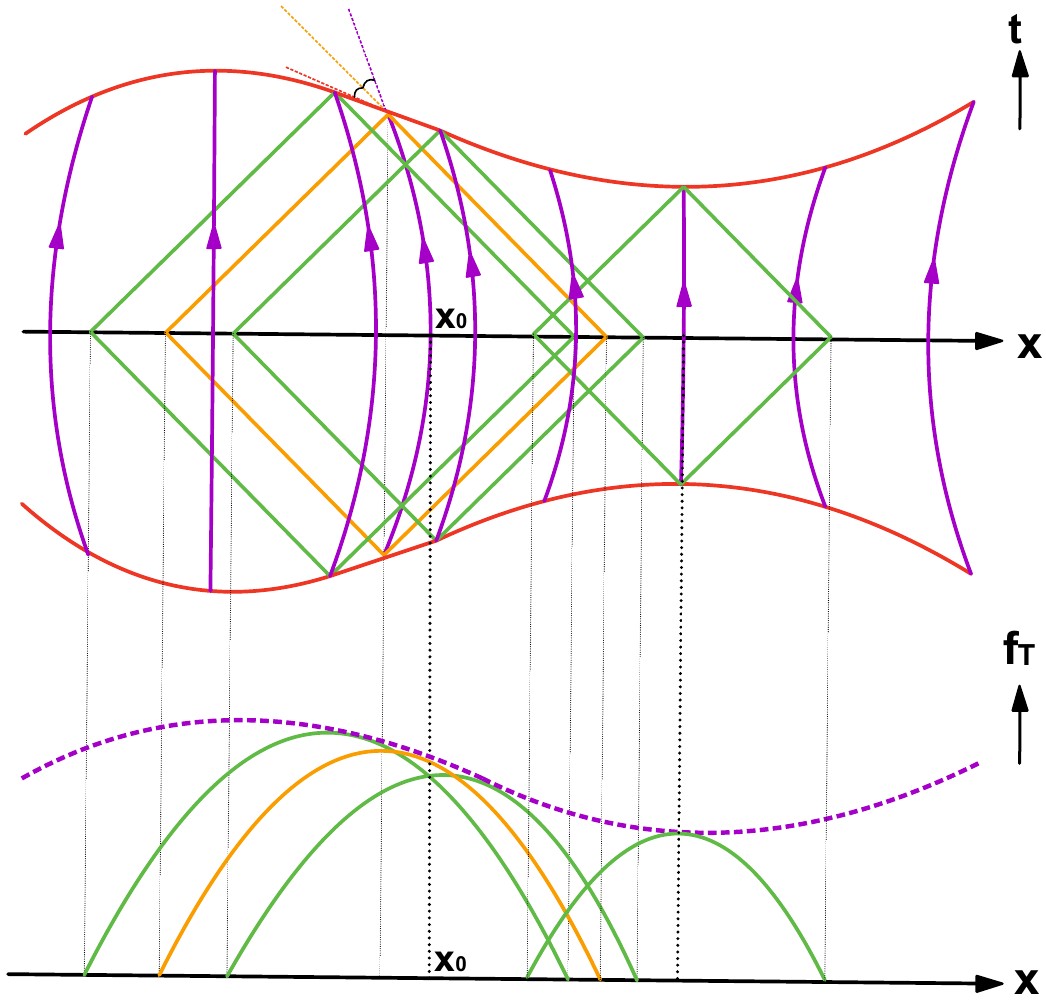}
 \caption{The upper figure shows the hyperbolic modular flow (observers' worldlines) in a general time band, and the lower figure depicts the entanglement temperature function $f_T(x)$ constructed by those observers, which is consistent with the $\max{\tilde{f}_{R_i}(x_0)}$ formula. For example, at point $x=x_0$ where the purple hyperbola inside the orange diamond intersects with the $x$ axis in the upper figure, the orange parabola in the lower figure happens to be the maximum among all other parabolas at $x=x_0$, and its value gives $f_T(x_0)$.}\label{hyperconsistent}
\end{figure}

\noindent
After the worldlines of observers are determined, we can write down the modular Hamiltonian {determined from the modular flows generated by these observers}. The rules are simple: observer $i$ whose worldline intersects with the $t=0$ Cauchy slice at $x=x_i$ observes an entanglement temperature $\tilde{f}_{R_i}(x)|_{x \to x_i}$ in their own causal diamond ($R_i$) as shown in Figure \ref{hyperconsistent}.  {This is due to the fact that the worldline of each observer is the same worldline in the corresponding causal diamond and therefore the entanglement temperature derived from the tangential vector $\xi^\mu$ of the worldline should be the same}. We can simply combine the entanglement temperature the corresponding observer observes at each spatial point to get the entire modular Hamiltonian that those observers detect in the whole spatial subregion.

Note that this method of deriving the modular Hamiltonian is inherently independent from the method in the last section. One might wonder if these two distinctly different methods yield the same modular Hamiltonian. The answer is precisely yes: the construction of {consistent geometric modular flows using }observers with hyperbolic worldlines indeed gives the maximum entanglement temperature result from the last section.
\begin{equation}\label{consist}
    \tilde{f}_{R_i}(x_i) = \max_j\{\tilde{f}_{R_j}(x_i)\},
\end{equation}
where $x_i$ is the $x$ coordinate where the hyperbola, which is orthogonal to the time band at the vertex of the causal diamond $R_i$, intersects the $x$ axis. The maximum on the right-hand side is with respect to all subregions whose vertexes of causal diamond $R_j$ touches the spacelike boundaries of the time band. The formula (\ref{consist}) points out that the entanglement temperature in the causal diamond that is particularly fixed by the constraint conditions above is precisely the largest entanglement temperature! 
The specific proof of (\ref{consist}) is presented in Appendix C. {The underlying physical reason for this remarkable coincidence is unknown, while it strongly indicates the physical consistency of the time band state.
}   

    \subsection{Modular Hamiltonians for different time bands}
\begin{figure}[H]
\centering
     \includegraphics[width=10.5cm]{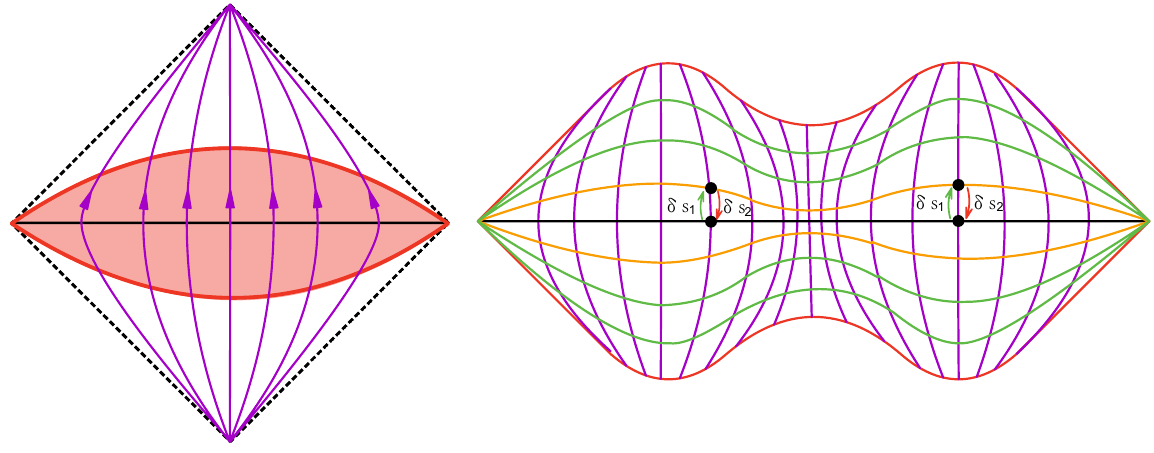}
 \caption{The left figure shows an ``eye-shaped" time band (time pancake in \cite{Witten:2018zxz}) whose modular flow coincides with the one in the entire causal diamond. The right figure illustrates the procedure in formula (\ref{Com}), evolving $\phi(x(0))$ via coinciding flows in different time bands.}\label{Commutation}
\end{figure}

\noindent We now further test the consistency of the modular Hamiltonians of different time bands from another aspect: the commutation relationship. {In the Tomita-Takesaki theory \cite{Haag1993LocalQP}}, the operators evolve along the modular flow as follows
\begin{equation}
    \phi(x(s)) = e^{-iKs} \phi(x(0)) e^{iKs}.
\end{equation}
Considering two time bands \(T_1\) and \(T_2\) with coinciding modular flows in their intersecting spacetime subregion (the smaller time band). This requires that the edges of these two time bands are orthogonal to the same set of hyperbolas.
As shown in Figure \ref{Commutation}, we can evolve an arbitrary operator $\phi(x(0))$ in the vacuum along the modular flow an infinitesimal distance $\delta s_1$ via the modular Hamiltonian of $T_1$, namely $K_1$, and then evolve it back to $\phi(x(0))$ along the modular flow of $T_2$ via $K_2$ at distance $\delta s_2$. After the evolving process, the operator should be the same operator as before at the same location
\begin{equation}\label{Com}
    \phi(x(0)) = e^{iK_2 \delta s_2} e^{-iK_1 \delta s_1} \phi(x(0)) e^{iK_1 \delta s_1} e^{-iK_2 \delta s_2},
\end{equation}
\ie, $e^{iK_2 \delta s_2} e^{-iK_1 \delta s_1}$ commutes with all operators in the vacuum state. Note that we cannot conclude that $e^{iK_2 \delta s_2} e^{-iK_1 \delta s_1}$ is a constant operator because only the operators in the vacuum state could not span a complete basis. However, this is still a very strong condition that constrains the formula of modular Hamiltonian. Consistently, as we proved in Appendix D, the modular Hamiltonian for the time band (\ref{consist}) naturally satisfies this condition, which confirms the consistency of the formula (\ref{consist}).

\section{Holographic entanglement first law}

\noindent One important application of the modular Hamiltonian is a consistency test for the entanglement first law \cite{Blanco:2013joa}. The entanglement first law states that {the variation in the entanglement entropy under  linear order perturbations in the state equals the first order variation in the expectation value of the modular Hamiltonian}

\begin{equation}
   \delta S(\rho)=\delta \left<H_\rho\right>.
\end{equation}

We study the entanglement first law for the holographic time band in the modified IR geometry via the Iyer-Wald formalism \cite{Iyer_1994}. We consider the simplest time band state with a uniform time cutoff for the whole spatial region. In the bulk, we modify the IR geometry by removing the $z>1$ region in Poincaré coordinates. In this case, the RT surface of the whole boundary system is simply the edge of the cut-off surface $z=1$. The bulk vector that generates the bulk modular flow
\begin{equation}
    \xi=\pi(1-z^2-t^2)\partial_{t}-2\pi t z\partial_{z}
    \label{xi}
\end{equation}
vanishes on the extremal surface $\tilde{B}$:$\{\left(t,x,z\right) |t=0,z=1\}$ and reduces to the generator of the boundary modular Hamiltonian on the asymptotic boundary $B$. $\xi$ is not a Killing vector while it satisfies the Killing equation on the $t=0$ hypersurface $\Sigma$.

We consider small perturbations around the Poincaré metric
\begin{equation}
    ds'^2=\frac{\eta_{\mu\nu}x^\mu dx^\nu+dz^2}{dz^2}+h_{\mu\nu}dx^\mu dx^\nu.
    \label{perturbation}
\end{equation}
To relate the boundary modular Hamiltonian and the area of bulk minimal surface, we introduce a bulk 1-form $\chi$ \cite{Faulkner_2014}
\begin{equation}
    \chi=\delta Q[\xi]-\xi \cdot \Theta[\xi],
\end{equation}
where $Q[\xi]$ is the Noether charge 1-form and  $\Theta[\xi]$ is the symplectic
 potential current 2-form \cite{Iyer_1994}. From (\ref{perturbation}) and (\ref{xi}), we can obtain the restriction of $\chi$ and $d\chi$ on $\Sigma$
\begin{equation}
    \chi|_{\Sigma}=\frac{h_{11}+z(1-z^2)\partial_z h_{11}}{8} dx,
\end{equation}
\begin{equation}
    d\chi|_{\Sigma}=\frac{(1-z^2)(3\partial_z h_{11}+z\partial^2_z h_{11})}{16} dx\wedge dz.
\end{equation}
One can check that $d\chi|_{\Sigma}=0$ from the $tt$-component of linearized Einstein equations. Therefore we have $\int_{B}\chi=\int_{\tilde{B}}\chi$ by Stokes theorem. The L.H.S is exactly the variation of the entanglement entropy
\begin{equation}
    \delta S=\frac{1}{4}\delta A=\frac{1}{8}\int dx \sqrt{g^{(0)}}g^{(0)11}h_{11}=\frac{1}{8}\int h_{11} dx=\int_{\tilde{B}}\chi.
\end{equation}
On the other hand
\begin{equation}
    \int_{B}\chi=\frac{1}{8}\int h_{11}(0,x,0) dx=\frac{1}{8}\int h_{00}(0,x,0)=\frac{1}{8}\int \left<T_{00}(0,x)\right>dx,
\end{equation}
and this is the variation of the modular Hamiltonian of the simplest time band state {$\delta \left<H\right>$}. The second equality comes from linearized Einstein equations, which also indicates the tracelessness of the CFT stress tensor. Thus we have verified $\delta S=\delta \left<H\right>$. More discussions on the entanglement first law in general time band states will be reported in \cite{Ju:2025}.

\section{Conclusion and discussion}
\noindent In this paper, we studied the form of the modular Hamiltonian for the holographic time band state. The time band is a causal incomplete subregion which corresponds to a causal complete convex subregion in the bulk. In a specific modified IR geometry, the bulk subregion ({causal wedge}) becomes an entanglement wedge. This suggests the existence of a local modular Hamiltonian for the holographic time band state. We have constructed the local modular Hamiltonian using two independent methods: from the quantum information properties of the time band state and from the construction of geometric modular flows. 
Both methods lead to the same unique result of the local modular Hamiltonian,
\begin{equation}\label{flatconclusion}
    f_T(x) = \max\{\tilde{f}_{R_i}(x)\},
\end{equation}
reflecting the intrinsic property of the time band state. The entanglement first law has been checked to hold for the simplest time band state.

Although our argument is formulated in AdS\(_3\)/CFT\(_2\), it can be easily generalized to higher dimensions because the formula (\ref{consist}) is universal. Both methods remain valid in higher dimensions.

{The local modular Hamiltonian derived in this work for the time band state will also help us solve the holographic causal information problem \cite{Bousso:2012sj,Hubeny:2012wa,Kelly_2014}, which involves finding the density matrix corresponding to the causal wedge. This is notably a problem when CW is different from the entanglement wedge, such as for a non-spherical boundary subregion in the vacuum state in higher dimensions. In this example, the spatial part of the causal wedge is bounded by a convex surface \cite{Balasubramanian:2013lsa,Hubeny:2013gba} in the bulk, and it becomes an entanglement wedge in an IR modified geometry. 
As a result, using the method presented in this work, one can construct a density matrix with a local modular Hamiltonian whose von Neumann entropy equals its area.
In this sense, the causal holographic information problem  in vacuum is solved.
}
Further discussion in detail will be seen in a forthcoming paper \cite{Ju:2025}.

\section*{Acknowledgement}
\noindent We thank Bart{\l}omiej Czech, Veronika Hubeny, Ling-Yan Hung, Tengzhou Lai, Hong Liu, Samuel Leutheusser, Wen-Bin Pan, Wei Song, Huajia Wang and Zhenbin Yang for useful discussions. This work was supported by Project 12035016 supported by the National Natural Science Foundation of China.

\appendix
\section{Uniqueness of the local modular Hamiltonian in flat case}
\noindent We denote the function of the entanglement temperature of the length $L+2R$ boundary subregion in the flat case $f_{T_{L+2R}}(x)$, which is a continuous (with respect to $L$ and $x$) and even function. Moreover, due to the vanishing condition of CMI (\ref{KCMI}), it must obey
\begin{equation}\label{Kcancel}
f_{T_{a+2R}}(x+\frac{a-c}{2}) + f_{T_{b+2R}}(x-\frac{b-c}{2}) - f_{T_{c+2R}}(x) - f_{T_{a+b-c+2R}}(x+\frac{a-b}{2}) = 0,
\end{equation}
where $R$ is the $z$ coordinate of the convex line which is parallel to the conformal boundary in Poincaré patch coordinates.
Utilizing this formula, we can find that once we get $f_{T_{L+2R}}(x)$ with $L>0$, we can use it to generate $f_{T_{2L+2R}}(x)$ with $a=b=L$ and $c=0$, then $f_{T_{4L+2R}}(x)$, etc. Moreover, we have
\begin{equation}\label{lim}
\begin{aligned}
    \lim_{L\to 0}f_{T_{L+2R}}(x)&=\pi\frac{R^2-x^2}{R}\\
    \lim_{L\to \infty}f_{T_{L+2R}}(x)&=\pi R, 
\end{aligned}
\end{equation}
where the $L\to \infty$ result is derived from \textbf{Constraint 3 and 4} in the main text.

It can be easily verified that the formula given in the paper
\begin{equation}\label{uniq}
    f_{T_{L+2R}}(x) = \max\{\tilde{f}_{R_i}(x)\} = \left\{
    \begin{aligned}
        &\pi\frac{R^2 - (x + L/2)^2}{R} \quad\quad\quad (x \in [-L/2 - R, -L/2]) \\
        &\pi R \quad\quad\quad\quad\quad\quad\quad\quad\quad (x \in [-L/2, L/2]) \\
        &\pi\frac{R^2 - (x - L/2)^2}{R} \quad\quad\quad (x \in [L/2, L/2 + R]), \\
    \end{aligned}
    \right.
\end{equation}
certainly fulfills all the conditions above. The question remains whether it is the unique function.
It can be assumed that \( f_{T_{L+2R}}(x) = \max\{\tilde{f}_{R_i}(x)\} + \lambda_{L+2R}(x) \), where $\lambda_{L+2R}(x)$ denotes possible deviation from the function above. Rewriting the constraints with respect to the \(\lambda\) function, we can find that the right-hand side of equation (\ref{lim}) becomes zero for the \(\lambda\) function, \ie, {
    $\lim_{L\to 0}\lambda_{T_{L+2R}}(x)=    \lim_{L\to \infty}\lambda_{T_{L+2R}}(x)=0. $}
 Due to \textbf{Constraint 4}, \(\lambda_{L+2R}(x) \geq 0\). The remaining work is to prove \(\lambda_{L+2R}(x) = 0\) for any \(L > 0\).

For a region whose length \(L+2R\) is greater than \(4R\), one can use (\ref{Kcancel}) twice (on the left side and the right side respectively) to generate \(\lambda_{3L+2R}(x)\):
\begin{equation}
    \lambda_{3L+2R}(x) = \lambda_{L+2R}(x) + \lambda_{L+2R}(x+L) + \lambda_{L+2R}(x-L).
\end{equation}
As \(\lambda_{L+2R}(x) = 0\) when \(|x| > L/2 + R\), we can derive that \(\lambda_{3L+2R}(x) = \lambda_{L+2R}(x)\) when \(|x| < L/2 - R\). Then, we can infinitely enlarge the boundary region so that we get \(\lim_{n \to \infty} \lambda_{(2n+1)L+2R}(x) = \lambda_{L+2R}(x) = 0\) due to equation (\ref{lim}).
To sum up, we have
\begin{equation}\label{zero}
    \lambda_{L+2R}(x) = 0 \quad\quad (L \geq 2R, \quad x \in [-L/2+R, L/2-R]).
\end{equation}
At the same time, we have
\begin{equation}
    \lambda_{L+2R}(x) = \lambda_{L+2R}(x-L/2) + \lambda_{L+2R}(x+L/2) = 0 \quad (L > 2R, \, -L/2+R < x < L/2-R)
\end{equation}
due to the positivity of the \(\lambda\) function. We have \(\lambda_{L+2R}(x \pm L/2) = 0\), where \(x \pm L/2\) covers all \(x \in [-L/2-R, L/2+R]\), which means \(\lambda_{L+2R}(x) = 0\) for \(L > 2R\).

We can then prove \(\lambda_{L+2R}(x) = 0\) for \(L \leq 2R\) because
\begin{equation}
    \lambda_{L+2R}(x) = \lambda_{4R}(x-R+L/2) + \lambda_{4R}(x+R-L/2) - \lambda_{6R-L}(x),
\end{equation}
where each term on the RHS equals zero. As a result, we have \(\lambda_{L+2R}(x) = 0\) for \(L \leq 2R\). To sum up, \(\lambda_{L+2R}(x) = 0\) for all positive \(L\), and the uniqueness of formula (\ref{uniq}) is proved.

\section{Non-intersecting condition of the worldlines of boundary observers}

\noindent In this appendix, we prove that the worldlines of boundary observers inside the time reflection symmetric time band generated by a convex curve in the bulk will never intersect, as long as the worldlines are:

\begin{itemize}
\item timelike hyperbolic flows with time reflection symmetry,
\item orthogonal to the time band.
\end{itemize}

First, we prove that the boundary of a time band generated by a convex surface must be ``convex" from the viewpoint of a spatial hyperbola (in Poincare coordinates); that is, any spatial hyperbola tangential to the boundary of the time band will not go through it. Note that the origin of this hyperbola must be on the Cauchy slice with time reflection symmetry.

The proof is quite straightforward. Given a point in the bulk, the light cone it emits will intersect the conformal boundary at two hyperbolas \cite{maldacena2015lookingbulkpoint}. In other words, the region causally disconnected from the point is a wedge intersecting the conformal boundary between these hyperbolas. Note that the time band generated by the convex region is the intersection of the regions causally disconnected from its boundary. As a result, a hyperbola tangential to the time band must be generated by a point on the convex curve, and the upper region of this hyperbola, causally connected with this point, must lie outside this time band.

\begin{figure}[H]
\centering
\includegraphics[width=10.5cm]{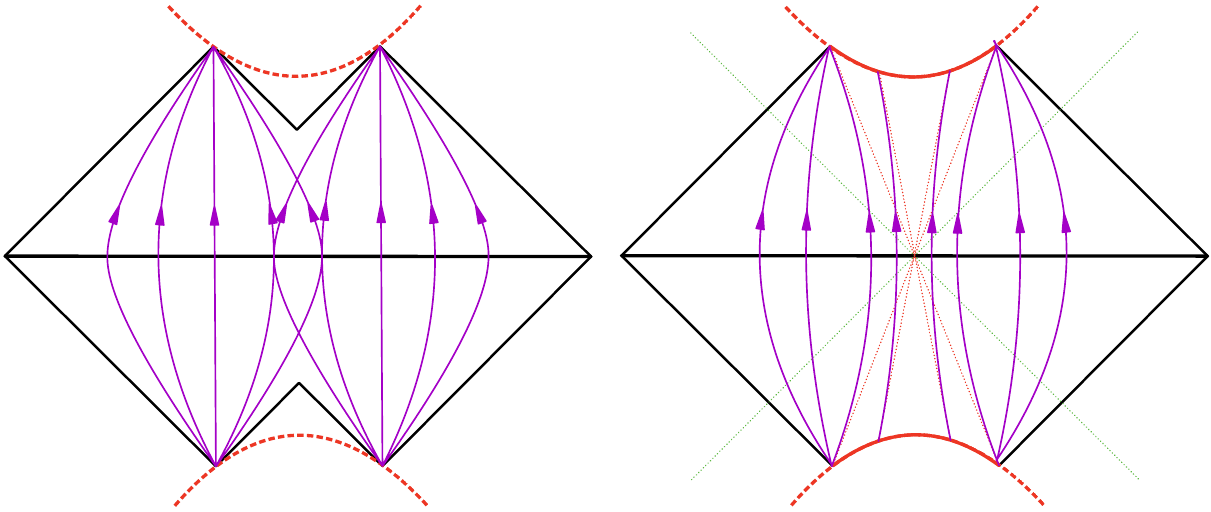}
\caption{The worldlines of observers intersect in a zigzag-shaped time band, which cannot be generated by a convex curve in the bulk (left), and do not intersect in a time band generated by a convex curve in the bulk (right).
}\label{nonintersect}
\end{figure}

As shown in Figure \ref{nonintersect}, a zigzag-shaped time band could not be generated by a convex curve in the bulk, while its ``hyperbola-convex hull," which is the region completed using hyperbolas tangential to it, can.

One can find in Figure \ref{nonintersect} that for the zigzag-shaped time band, the hyperbolas generated via the two conditions above can actually intersect with each other. This poses a serious problem in formulating the modular Hamiltonian, as there will exist more than one observer who observes different marginal states at the same point, making the entanglement temperature not unique at this point, and rendering the procedure inconsistent by definition.

However, we can prove that this will never happen for a time band generated by a convex curve in the bulk. To show this, we can draw lines tangential to the hyperbola on the boundary of the time band, shown as red dotted lines on the right in Figure \ref{nonintersect}. The necessary condition for two hyperbolas to intersect when $t>0$ is that these lines intersect when $t>0$. However, these lines cannot intersect when $t>0$ as follows.

Due to the geometrical properties of hyperbolas, the line tangential to the worldline of observers on the boundary of the time band is the line passing through the origin of the hyperbola orthogonal to it, \eg, the boundary of the time band. For the lines to intersect when $t > 0$, the boundary of the time band must be wiggly. However, the boundary of the time band is ``hyperbolic-convex," which is most wiggly when it is partially hyperbolic. In this case, the lines intersect at the origin of the hyperbola at $t = 0$ rather than $t > 0$. As a result, these lines never intersect, which leads to the conclusion that the worldlines of observers in the time band will never intersect as long as the time band is generated by a convex curve in the bulk.

\section{Equivalence of two formulas of modular Hamiltonian}
\noindent In this appendix, we are going to prove the equivalence of the formulas for the modular Hamiltonian obtained via two different methods, {\ie, formula (\ref{consist})}. These two distinct methods are the ``maximizing method," where the entanglement temperature of {the time band state}is the maximum value of the entanglement temperature of all marginal states, and the {consistent modular flow method}, where the entanglement temperature of {the time band state} is the entanglement temperature each observer (with specific hyperbolic worldlines) observes in its own causal diamond.

To prove their equivalence, we must show that the entanglement temperature (which observers detect at their location within their own causal diamond) is the maximum value among all entanglement temperatures at this point of any marginal state, as shown in Figure \ref{hyperconsistent}.

Given the (orange) causal diamond an observer resides in is of length \( 2R(x_0) \), and the (green) diamond next to it is \( 2R(x_0) + 2R'(x_0)dx \), with the derivative of the \( R \) function being \( R'(x_0) \), the equation of the hyperbola, which is the worldline of the observer, is
\begin{equation}
    (x - x_0 + \frac{R(x_0)}{R'(x_0)})^2 - y^2 = R(x_0)^2 \left(\frac{1}{R'(x_0)^2} - 1\right),
\end{equation}
One can check that it is orthogonal to the boundary of the time band at \( x = x_0 \), and it intersects with \( t = 0 \) at the location
\begin{equation}\label{hypflowpoint}
    x = x_0 - R(x_0) \frac{1 - \sqrt{1 - R'(x_0)^2}}{R'(x_0)}.
\end{equation}

On the other hand, the function of entanglement temperature of {the time band state} via the maximizing method near \( x_0 \) is 
\begin{equation}
    f_{T}(x) =\pi \max \left\{ \frac{(R(x_0) + R'(x_0)x_1)^2 - (x - x_0 - x_1)^2}{R(x_0) + R'(x_0)x_1} \right\},
\end{equation}
where \( x_1 \) is a small value labeling the causal diamond near \( x = x_0 \).
To get the maximum value among all \( x_1 \) chosen, one has to solve
\begin{equation}
    \partial_{x_1} f_{T}(x_0) = 0
\end{equation}
for \( x_1 \) to determine which diamond reaches the maximum value of entanglement temperature at \( x = x_0 \). The solution is 
\begin{equation}
    x_1 = \frac{R(x_0)R'(x_0)}{1 - R'(x_0)^2 + \sqrt{1 - R'(x_0)^2}}.
\end{equation}
Next, we calculate the \( x \) value for \( f_{R(x_0)}(x) \) that reaches the maximum among all other marginal states. This can be easily calculated using the homothety relationship, and the result is given by
\begin{equation}
    x = x_0 - \frac{R(x_0)x_1}{R(x_0) + R'(x_0)x_1} = x_0 - R(x_0) \frac{1 - \sqrt{1 - R'(x_0)^2}}{R'(x_0)}.
\end{equation}
To sum up, the causal diamond at \( x = x_0 \) reaches the maximum among all \( f \) functions of marginal states at this point, which exactly matches with equation (\ref{hypflowpoint}), meaning the two formulas for the modular Hamiltonian are equivalent.

\section{Commutation of modular Hamiltonians with same modular flow}
\noindent In this appendix, we prove that the entanglement temperature function of different time bands with partially coinciding modular flows is proportional to each other.

We start with the simplest case, the ``eye-shaped" time band, as shown on the left side of Figure \ref{Commutation}. The worldlines of observers are the vertical hyperbolas intersecting at the upper and lower tips of the diamond. The horizontal hyperbolas denote constant time slices with respect to those observers. In this case, the formula for the coordinate transformation to the observers' reference frame is given by:
\begin{equation}
    X = \frac{(t^2 - 1)}{t^2 x^2 - 1}x, \quad\quad
    T = \frac{(x^2 - 1)}{t^2 x^2 - 1}t
\end{equation}
where $X, T$ are the Minkowski coordinates, and $x, t$ are the spatial and time coordinates of the observers, which coincide with $X, T$ when $x = 0$ or $t = 0$. The curves with constant $x$ and $t$ are spacelike and timelike hyperbolas, respectively, and the radius of the entire causal diamond is set to one ($R = 1$).

For an observer $O_1$ in this ``eye-shaped" time band $T_1$ with the edge of this time band a time slice with constant $t$ coordinate, the entanglement temperature it observes is
\begin{equation}
    f_T(x) = \pi \frac{\left(\frac{(x^2 - 1)}{t^2 x^2 - 1} t\right)^2 - \left(x-\frac{(t^2 - 1)}{t^2 x^2 - 1} x\right)^2}{\frac{(x^2 - 1)}{t^2 x^2 - 1} t} = \pi t (1 - x^2),
\end{equation}
which is exactly proportional to the entanglement temperature function of the entire causal diamond, e.g., $f_{R=1}(x) = \pi (1 - x^2)$.

\begin{figure}[H]
\centering
     \includegraphics[width=10.5cm]{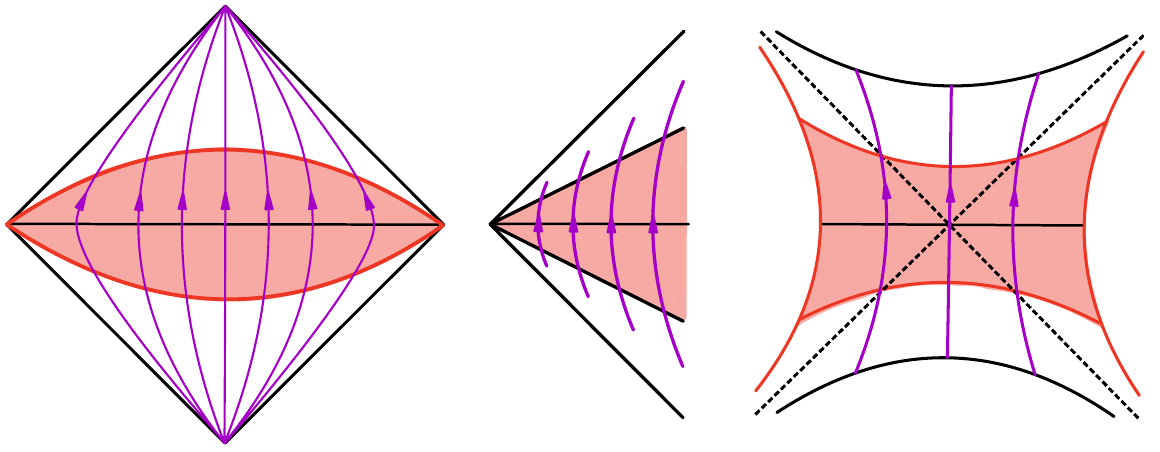}
 \caption{Eye-shaped time band (left), Rindler-shaped time band (middle), and hyperbolic-shaped time band (right). The red shaded regions are their respective sub-time bands, which have the coinciding modular flow marked as purple arrow curves.
}\label{tricases}
\end{figure}

A similar result can be proved in the Rindler-shaped time band case and the hyperbolic-shaped time band case as shown in Figure \ref{tricases}. Note that the Rindler-shaped time band can be regarded as an infinitely large eye-shaped time band, so its sub-time band with coinciding modular flow must have an entanglement temperature function proportional to its entanglement temperature function, as described above. In the hyperbolic-shaped time band, the coordinate transformation is given by
\begin{equation}
    X = \frac{(t^2 + 1)}{1 - t^2 x^2} x, \quad\quad
    T = \frac{(x^2 + 1)}{1 - t^2 x^2} t.
\end{equation}
For a single observer in this hyperbolic-shaped region, the entanglement temperature it detects within a time band with constant $t$ is
\begin{equation}
    f_T(x) = \pi \frac{\left(\frac{(x^2 + 1)}{1 - t^2 x^2} t\right)^2 - \left(x-\frac{(t^2 + 1)}{1 - t^2 x^2} x\right)^2}{\frac{(x^2 + 1)}{1 - t^2 x^2} t} = \pi t (1 + x^2).
\end{equation}
In fact, we can see that the hyperbolic time band case is a rotation in the complex axis of the ``eye-shaped" time band case, and it indeed has a proportional entanglement temperature function with respect to the coefficient $t$.

\begin{figure}[H]
\centering
     \includegraphics[width=10.5cm]{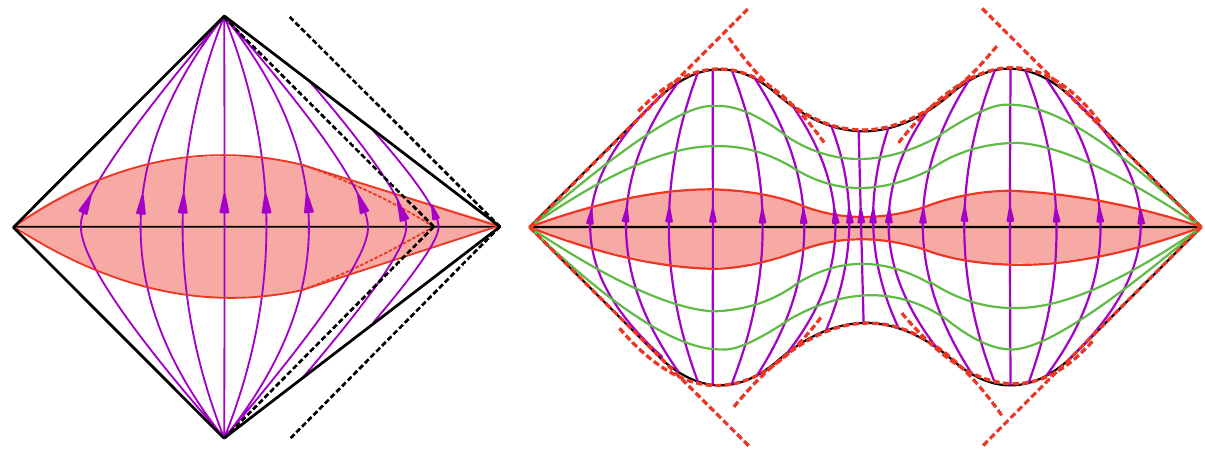}
 \caption{Left: a combination of Rindler-shaped time band and eye-shaped time band, which is smoothly connected. Right: an arbitrary time band generated by a convex curve in the bulk could be interpolated using the three small types of time bands above, marked by red dashed curves, which are smoothly connected, and their sub-time bands are smoothly connected as well.}\label{interpolate}
\end{figure}

Now we consider a time band whose boundary is a smooth combination of two of the three curves, and investigate the entanglement temperature function of it.
As shown in the Figure \ref{interpolate} as an example, smoothly combining the ``Rindler-shaped time band" and ``eye-shaped" time band together into a large time band, due to geometrical properties, its sub-time band with coinciding modular flow is also a time band smoothly connected by a ``Rindler-shaped {sub}-time band" and ``eye-shaped" sub-time band. 

Now, investigating their entanglement temperature function, due to the ``observer formula", it must be a function smoothly connected by the two entanglement temperature functions of the ``Rindler-shaped time band" and the ``eye-shaped" time band. As long as it is smoothly connected, and considering the fact that the time band on each side has a proportional entanglement temperature function, their combination must be proportional as well. Similarly, one can prove that the combination of the boundary of the time band using the three curves mentioned above must have a proportional entanglement temperature of its sub-time band with coinciding modular flow.

Given an arbitrary time band generated by a convex curve in the bulk, we can use the second-order interpolation function method, combining small intervals using those three curves smoothly together to approximate the real boundary of the time band. Specifically, when the second-order derivative of the boundary of the time band is positive, we can interpolate it using small hyperbolic-shaped time bands. When it is zero, we can interpolate it using small Rindler-shaped time bands, and when it is negative, we can use small eye-shaped time bands. This is why we consider those three specific shapes of time bands in the first place.

We have learned that any time band whose boundary is the combination of those three curves must have an exactly proportional entanglement temperature behavior. Now, we know that any time band can be regarded as an infinite combination of those three curves using the interpolation function method. As a result, we can conclude that any time band generated by a convex curve in the bulk must have an exactly proportional entanglement temperature behavior.

Finally, the coefficient of the proportional entanglement temperature at $x = x_0$ is checked to be the ratio of two conformal Killing vectors at $x = x_0$ inside the respective causal diamonds. As a result, we have
\begin{equation}
    K_1 \delta s_1 = K_2 \delta s_2 + C,
\end{equation}
where $s_1$ and $s_2$ are the "times" of these evolutions, which are inversely proportional to the conformal Killing vectors at $x = x_0$ inside the respective causal diamonds. 

End proof.

\bibliography{reference}

\providecommand{\href}[2]{#2}\begingroup\raggedright\begin{thebibliography}{10}

\bibitem{Maldacena:1997re}
J.~M. Maldacena, \emph{{The Large N limit of superconformal field theories and supergravity}}, \href{http://dx.doi.org/10.4310/ATMP.1998.v2.n2.a1}{\emph{Adv. Theor. Math. Phys.} {\bf 2} (1998) 231--252}, [\href{http://arxiv.org/abs/hep-th/9711200}{{\tt hep-th/9711200}}].

\bibitem{Ryu:2006bv}
S.~Ryu and T.~Takayanagi, \emph{{Holographic derivation of entanglement entropy from AdS/CFT}}, \href{http://dx.doi.org/10.1103/PhysRevLett.96.181602}{\emph{Phys. Rev. Lett.} {\bf 96} (2006) 181602}, [\href{http://arxiv.org/abs/hep-th/0603001}{{\tt hep-th/0603001}}].

\bibitem{Ryu:2006ef}
S.~Ryu and T.~Takayanagi, \emph{{Aspects of Holographic Entanglement Entropy}}, \href{http://dx.doi.org/10.1088/1126-6708/2006/08/045}{\emph{JHEP} {\bf 08} (2006) 045}, [\href{http://arxiv.org/abs/hep-th/0605073}{{\tt hep-th/0605073}}].

\bibitem{Czech:2012bh}
B.~Czech, J.~L. Karczmarek, F.~Nogueira and M.~Van~Raamsdonk, \emph{{The Gravity Dual of a Density Matrix}}, \href{http://dx.doi.org/10.1088/0264-9381/29/15/155009}{\emph{Class. Quant. Grav.} {\bf 29} (2012) 155009}, [\href{http://arxiv.org/abs/1204.1330}{{\tt 1204.1330}}].

\bibitem{Wall:2012uf}
A.~C. Wall, \emph{{Maximin Surfaces, and the Strong Subadditivity of the Covariant Holographic Entanglement Entropy}}, \href{http://dx.doi.org/10.1088/0264-9381/31/22/225007}{\emph{Class. Quant. Grav.} {\bf 31} (2014) 225007}, [\href{http://arxiv.org/abs/1211.3494}{{\tt 1211.3494}}].

\bibitem{Headrick:2014cta}
M.~Headrick, V.~E. Hubeny, A.~Lawrence and M.~Rangamani, \emph{{Causality \& holographic entanglement entropy}}, \href{http://dx.doi.org/10.1007/JHEP12(2014)162}{\emph{JHEP} {\bf 12} (2014) 162}, [\href{http://arxiv.org/abs/1408.6300}{{\tt 1408.6300}}].

\bibitem{Bousso:2022hlz}
R.~Bousso and G.~Penington, \emph{{Entanglement Wedges for Gravitating Regions}},  \href{http://arxiv.org/abs/2208.04993}{{\tt 2208.04993}}.

\bibitem{Espindola:2018ozt}
R.~Esp\'\i{}ndola, A.~Guijosa and J.~F. Pedraza, \emph{{Entanglement Wedge Reconstruction and Entanglement of Purification}}, \href{http://dx.doi.org/10.1140/epjc/s10052-018-6140-2}{\emph{Eur. Phys. J. C} {\bf 78} (2018) 646}, [\href{http://arxiv.org/abs/1804.05855}{{\tt 1804.05855}}].

\bibitem{Saraswat:2020zzf}
K.~Saraswat and N.~Afshordi, \emph{{Extracting Hawking Radiation Near the Horizon of AdS Black Holes}}, \href{http://dx.doi.org/10.1007/JHEP02(2021)077}{\emph{JHEP} {\bf 02} (2021) 077}, [\href{http://arxiv.org/abs/2003.12676}{{\tt 2003.12676}}].

\bibitem{dong2016reconstruction}
X.~Dong, D.~Harlow and A.~C. Wall, \emph{Reconstruction of bulk operators within the entanglement wedge in gauge-gravity duality}, {\emph{Physical review letters} {\bf 117} (2016) 021601}.

\bibitem{Harlow:2018fse}
D.~Harlow, \emph{{TASI Lectures on the Emergence of Bulk Physics in AdS/CFT}}, \href{http://dx.doi.org/10.22323/1.305.0002}{\emph{PoS} {\bf TASI2017} (2018) 002}, [\href{http://arxiv.org/abs/1802.01040}{{\tt 1802.01040}}].

\bibitem{Bousso:2012sj}
R.~Bousso, S.~Leichenauer and V.~Rosenhaus, \emph{{Light-sheets and AdS/CFT}}, \href{http://dx.doi.org/10.1103/PhysRevD.86.046009}{\emph{Phys. Rev. D} {\bf 86} (2012) 046009}, [\href{http://arxiv.org/abs/1203.6619}{{\tt 1203.6619}}].

\bibitem{Leutheusser:2022bgi}
S.~Leutheusser and H.~Liu, \emph{{Subalgebra-subregion duality: emergence of space and time in holography}},  \href{http://arxiv.org/abs/2212.13266}{{\tt 2212.13266}}.

\bibitem{Balasubramanian:2013rqa}
V.~Balasubramanian, B.~Czech, B.~D. Chowdhury and J.~de~Boer, \emph{{The entropy of a hole in spacetime}}, \href{http://dx.doi.org/10.1007/JHEP10(2013)220}{\emph{JHEP} {\bf 10} (2013) 220}, [\href{http://arxiv.org/abs/1305.0856}{{\tt 1305.0856}}].

\bibitem{Balasubramanian:2013lsa}
V.~Balasubramanian, B.~D. Chowdhury, B.~Czech, J.~de~Boer and M.~P. Heller, \emph{{Bulk curves from boundary data in holography}}, \href{http://dx.doi.org/10.1103/PhysRevD.89.086004}{\emph{Phys. Rev. D} {\bf 89} (2014) 086004}, [\href{http://arxiv.org/abs/1310.4204}{{\tt 1310.4204}}].

\bibitem{Hubeny_2014}
V.~E. Hubeny, \emph{Covariant residual entropy}, \href{http://dx.doi.org/10.1007/jhep09(2014)156}{\emph{Journal of High Energy Physics} {\bf 2014} (sep, 2014) }.

\bibitem{Ju:2023bjl}
X.-X. Ju, W.-B. Pan, Y.-W. Sun and Y.-T. Wang, \emph{{Generalized Rindler Wedge and Holographic Observer Concordance}},  \href{http://arxiv.org/abs/2302.03340}{{\tt 2302.03340}}.

\bibitem{Ju:2023dzo}
X.-X. Ju, B.-H. Liu, W.-B. Pan, Y.-W. Sun and Y.-T. Wang, \emph{{Squashed Entanglement from Generalized Rindler Wedge}},  \href{http://arxiv.org/abs/2310.09799}{{\tt 2310.09799}}.

\bibitem{Czech:2014tva}
B.~Czech, P.~Hayden, N.~Lashkari and B.~Swingle, \emph{{The Information Theoretic Interpretation of the Length of a Curve}}, \href{http://dx.doi.org/10.1007/JHEP06(2015)157}{\emph{JHEP} {\bf 06} (2015) 157}, [\href{http://arxiv.org/abs/1410.1540}{{\tt 1410.1540}}].

\bibitem{Balasubramanian:2018uus}
V.~Balasubramanian and C.~Rabideau, \emph{{The dual of non-extremal area: differential entropy in higher dimensions}}, \href{http://dx.doi.org/10.1007/JHEP09(2020)051}{\emph{JHEP} {\bf 09} (2020) 051}, [\href{http://arxiv.org/abs/1812.06985}{{\tt 1812.06985}}].

\bibitem{Hubeny:2014qwa}
V.~E. Hubeny, \emph{{Covariant Residual Entropy}}, \href{http://dx.doi.org/10.1007/JHEP09(2014)156}{\emph{JHEP} {\bf 09} (2014) 156}, [\href{http://arxiv.org/abs/1406.4611}{{\tt 1406.4611}}].

\bibitem{Headrick:2014eia}
M.~Headrick, R.~C. Myers and J.~Wien, \emph{{Holographic Holes and Differential Entropy}}, \href{http://dx.doi.org/10.1007/JHEP10(2014)149}{\emph{JHEP} {\bf 10} (2014) 149}, [\href{http://arxiv.org/abs/1408.4770}{{\tt 1408.4770}}].

\bibitem{Myers:2014jia}
R.~C. Myers, J.~Rao and S.~Sugishita, \emph{{Holographic Holes in Higher Dimensions}}, \href{http://dx.doi.org/10.1007/JHEP06(2014)044}{\emph{JHEP} {\bf 06} (2014) 044}, [\href{http://arxiv.org/abs/1403.3416}{{\tt 1403.3416}}].

\bibitem{Czech:2014wka}
B.~Czech, X.~Dong and J.~Sully, \emph{{Holographic Reconstruction of General Bulk Surfaces}}, \href{http://dx.doi.org/10.1007/JHEP11(2014)015}{\emph{JHEP} {\bf 11} (2014) 015}, [\href{http://arxiv.org/abs/1406.4889}{{\tt 1406.4889}}].

\bibitem{Hayden:2004wwj}
P.~Hayden, R.~Jozsa, D.~Petz and A.~Winter, \emph{{Structure of States Which Satisfy Strong Subadditivity of Quantum Entropy with Equality}}, \href{http://dx.doi.org/10.1007/s00220-004-1049-z}{\emph{Commun. Math. Phys.} {\bf 246} (2004) 359--374}.

\bibitem{Fawzi_2015}
O.~Fawzi and R.~Renner, \emph{Quantum conditional mutual information and approximate markov chains}, \href{http://dx.doi.org/10.1007/s00220-015-2466-x}{\emph{Communications in Mathematical Physics} {\bf 340} (Sept., 2015) 575–611}.

\bibitem{Ju:2024xcn}
X.-X. Ju, T.-Z. Lai, B.-H. Liu, W.-B. Pan and Y.-W. Sun, \emph{{Entanglement structures from modified IR geometry}},  \href{http://arxiv.org/abs/2404.02737}{{\tt 2404.02737}}.

\bibitem{Chen:2022eyi}
L.~Chen and H.~Wang, \emph{{Causal shadow and non-local modular flow: from degeneracy to perturbative genesis by correlation}}, \href{http://dx.doi.org/10.1007/JHEP02(2023)052}{\emph{JHEP} {\bf 02} (2023) 052}, [\href{http://arxiv.org/abs/2211.12064}{{\tt 2211.12064}}].

\bibitem{Haag1993LocalQP}
R.~Haag, \emph{Local quantum physics : fields, particles, algebras},  1993.

\bibitem{Araki:1976zv}
H.~Araki, \emph{{Relative Entropy of States of Von Neumann Algebras}}, {\emph{Publ. Res. Inst. Math. Sci. Kyoto} {\bf 1976} (1976) 809--833}.

\bibitem{Casini:2008cr}
H.~Casini, \emph{{Relative entropy and the Bekenstein bound}}, \href{http://dx.doi.org/10.1088/0264-9381/25/20/205021}{\emph{Class. Quant. Grav.} {\bf 25} (2008) 205021}, [\href{http://arxiv.org/abs/0804.2182}{{\tt 0804.2182}}].

\bibitem{Jafferis:2015del}
D.~L. Jafferis, A.~Lewkowycz, J.~Maldacena and S.~J. Suh, \emph{{Relative entropy equals bulk relative entropy}}, \href{http://dx.doi.org/10.1007/JHEP06(2016)004}{\emph{JHEP} {\bf 06} (2016) 004}, [\href{http://arxiv.org/abs/1512.06431}{{\tt 1512.06431}}].

\bibitem{Blanco:2013joa}
D.~D. Blanco, H.~Casini, L.-Y. Hung and R.~C. Myers, \emph{{Relative Entropy and Holography}}, \href{http://dx.doi.org/10.1007/JHEP08(2013)060}{\emph{JHEP} {\bf 08} (2013) 060}, [\href{http://arxiv.org/abs/1305.3182}{{\tt 1305.3182}}].

\bibitem{Hubeny:2007xt}
V.~E. Hubeny, M.~Rangamani and T.~Takayanagi, \emph{{A Covariant holographic entanglement entropy proposal}}, \href{http://dx.doi.org/10.1088/1126-6708/2007/07/062}{\emph{JHEP} {\bf 07} (2007) 062}, [\href{http://arxiv.org/abs/0705.0016}{{\tt 0705.0016}}].

\bibitem{Casini:2011kv}
H.~Casini, M.~Huerta and R.~C. Myers, \emph{{Towards a derivation of holographic entanglement entropy}}, \href{http://dx.doi.org/10.1007/JHEP05(2011)036}{\emph{JHEP} {\bf 05} (2011) 036}, [\href{http://arxiv.org/abs/1102.0440}{{\tt 1102.0440}}].

\bibitem{Bisognano-Wichmann}
J.~J. Bisognano and E.~H. Wichmann, \emph{On the duality condition for a hermitian scalar field}, \href{http://dx.doi.org/10.1063/1.522605}{\emph{Journal of Mathematical Physics} {\bf 16} (04, 1975) 985--1007}, [\href{http://arxiv.org/abs/https://pubs.aip.org/aip/jmp/article-pdf/16/4/985/19024093/985\_1\_online.pdf}{{\tt https://pubs.aip.org/aip/jmp/article-pdf/16/4/985/19024093/985\_1\_online.pdf}}].

\bibitem{Hislop:1981uh}
P.~D. Hislop and R.~Longo, \emph{{Modular Structure of the Local Algebras Associated With the Free Massless Scalar Field Theory}}, \href{http://dx.doi.org/10.1007/BF01208372}{\emph{Commun. Math. Phys.} {\bf 84} (1982) 71}.

\bibitem{Jensen:2023yxy}
K.~Jensen, J.~Sorce and A.~J. Speranza, \emph{{Generalized entropy for general subregions in quantum gravity}}, \href{http://dx.doi.org/10.1007/JHEP12(2023)020}{\emph{JHEP} {\bf 12} (2023) 020}, [\href{http://arxiv.org/abs/2306.01837}{{\tt 2306.01837}}].

\bibitem{Chandrasekaran:2022cip}
V.~Chandrasekaran, R.~Longo, G.~Penington and E.~Witten, \emph{{An algebra of observables for de Sitter space}}, \href{http://dx.doi.org/10.1007/JHEP02(2023)082}{\emph{JHEP} {\bf 02} (2023) 082}, [\href{http://arxiv.org/abs/2206.10780}{{\tt 2206.10780}}].

\bibitem{Hubeny:2012wa}
V.~E. Hubeny and M.~Rangamani, \emph{{Causal Holographic Information}}, \href{http://dx.doi.org/10.1007/JHEP06(2012)114}{\emph{JHEP} {\bf 06} (2012) 114}, [\href{http://arxiv.org/abs/1204.1698}{{\tt 1204.1698}}].

\bibitem{Kelly_2014}
W.~R. Kelly and A.~C. Wall, \emph{Coarse-grained entropy and causal holographic information in ads/cft}, \href{http://dx.doi.org/10.1007/jhep03(2014)118}{\emph{Journal of High Energy Physics} {\bf 2014} (Mar., 2014) }.

\bibitem{Ju:2025}
X.-X. Ju, B.-H. Liu, Y.-W. Sun and Y.~Zhao, \emph{{In progress}}, .

\bibitem{Hayden_2004}
P.~Hayden, R.~Jozsa, D.~Petz and A.~Winter, \emph{Structure of states which satisfy strong subadditivity of quantum entropy with equality}, \href{http://dx.doi.org/10.1007/s00220-004-1049-z}{\emph{Communications in Mathematical Physics} {\bf 246} (Apr., 2004) 359–374}.

\bibitem{Ruskai_2002}
M.~B. Ruskai, \emph{Inequalities for quantum entropy: A review with conditions for equality}, \href{http://dx.doi.org/10.1063/1.1497701}{\emph{Journal of Mathematical Physics} {\bf 43} (Sept., 2002) 4358–4375}.

\bibitem{Lieb:1973cp}
E.~H. Lieb and M.~B. Ruskai, \emph{{Proof of the strong subadditivity of quantum-mechanical entropy}}, \href{http://dx.doi.org/10.1063/1.1666274}{\emph{J. Math. Phys.} {\bf 14} (1973) 1938--1941}.

\bibitem{Witten:2018zxz}
E.~Witten, \emph{{APS Medal for Exceptional Achievement in Research: Invited article on entanglement properties of quantum field theory}}, \href{http://dx.doi.org/10.1103/RevModPhys.90.045003}{\emph{Rev. Mod. Phys.} {\bf 90} (2018) 045003}, [\href{http://arxiv.org/abs/1803.04993}{{\tt 1803.04993}}].

\bibitem{Iyer_1994}
V.~Iyer and R.~M. Wald, \emph{Some properties of the noether charge and a proposal for dynamical black hole entropy}, \href{http://dx.doi.org/10.1103/physrevd.50.846}{\emph{Physical Review D} {\bf 50} (July, 1994) 846–864}.

\bibitem{Faulkner_2014}
T.~Faulkner, M.~Guica, T.~Hartman, R.~C. Myers and M.~Van~Raamsdonk, \emph{Gravitation from entanglement in holographic cfts}, \href{http://dx.doi.org/10.1007/jhep03(2014)051}{\emph{Journal of High Energy Physics} {\bf 2014} (Mar., 2014) }.

\bibitem{Hubeny:2013gba}
V.~E. Hubeny, M.~Rangamani and E.~Tonni, \emph{{Global properties of causal wedges in asymptotically AdS spacetimes}}, \href{http://dx.doi.org/10.1007/JHEP10(2013)059}{\emph{JHEP} {\bf 10} (2013) 059}, [\href{http://arxiv.org/abs/1306.4324}{{\tt 1306.4324}}].

\bibitem{maldacena2015lookingbulkpoint}
J.~Maldacena, D.~Simmons-Duffin and A.~Zhiboedov, \emph{Looking for a bulk point},  2015.

\end{thebibliography}\endgroup
\bibliographystyle{JHEP}
\end{document}